\documentclass{elsart}

\usepackage{graphicx}

\begin{document}

\bibliographystyle{prsty}

\begin{frontmatter}
\title{Hard photon and neutral pion production in cold nuclear matter}

\author[GANIL]{L.~Aphecetche\thanksref{SUBATECH}},
\author[KVI]{J.~Bacelar},
\author[GANIL]{H.~Delagrange\thanksref{SUBATECH}},
\author[GANIL]{D.~d'Enterria\thanksref{SUBATECH}},
\author[KVI]{M.~Hoefman},
\author[KVI]{H.~Huisman},
\author[KVI]{N.~Kalantar-Nayestanaki},
\author[KVI]{H.~L\"ohner},
\author[GANIL]{G.~Mart\'{\i}nez\thanksref{SUBATECH}},
\author[WARSAW]{T.~Matulewicz},
\author[KVI]{J.~Messchendorp},
\author[GANIL]{M.-J.~Mora\thanksref{SUBATECH}}, 
\author[KVI]{R.~Ostendorf},
\author[KVI]{S.~Schadmand\thanksref{GIESSEN}},
\author[GANIL]{Y.~Schutz\thanksref{SUBATECH}},
\author[KVI]{M.~Seip},
\author[REZ]{A.~Taranenko}, 
\author[GANIL]{R.~Turrisi\thanksref{INFN}},
\author[KVI]{M.-J.~Van Goethem\thanksref{NSCL}},
\author[KVI]{M.~Volkerts},
\author[REZ]{V.~Wagner},
\author[KVI]{H.W.~Wilschut}

\address[GANIL]{Grand Acc\'el\'erateur National d'Ions Lourds, 
          IN2P3-CNRS, DSM-CEA, BP~5027, 14076 Caen Cedex 5, France}
\address[KVI]{Kernfysisch Versneller Instituut, NL-9747 AA Groningen, 
          The Netherlands}
\address[WARSAW]{Institute of Experimental Physics, Warsaw University,
        PL-00681 Warsaw, Poland}
\address[REZ]{Institute of Nuclear Physics, CZ-25068 \v Re\v z, 
             Czech Republic}

\thanks[SUBATECH]{Present address: SUBATECH, 4 rue A. Kastler, 
        F-44307 Nantes, France}
\thanks[GIESSEN]{Present address: II. Physikalisches Institut, 
        Universit\"at Gie\ss en, D-35392 Gie\ss en, Germany} 
\thanks[INFN]{Present address: INFN-Padova, Via Marzolo 8, 35131 Padova, Italy}
\thanks[NSCL]{Present address: NSCL, MSU, East Lansing, 
        Michigan 48824-1321. USA.} 


\begin{abstract}
The production of hard photons and neutral pions in 190 MeV proton induced 
reactions on C, Ca, Ni, and W targets has been for the first time concurrently 
studied.
Angular distributions and energy spectra up to the kinematical limit 
are discussed and the production cross-sections are presented. 
From the target mass dependence of the cross-sections the propagation of pions 
through nuclear matter is analyzed and the production mechanisms of 
hard photons and primordial pions are derived. 
It is found that the production of subthreshold particles 
proceeds mainly through first chance nucleon-nucleon collisions. For the 
most energetic particles the mass scaling evidences 
the effect of multiple collisions.
\end{abstract}


\end{frontmatter}

\section{Introduction}

Forming hot and compressed nuclear matter by colliding heavy ions provides 
the today only known experimental mean to explore the phase diagram of 
nuclear matter.
During the collision the longitudinal momentum of the projectile is converted
by multiple nucleon-nucleon collisions into transverse momentum as well as 
through the production of secondary particles, leading to
the creation of a hot and compressed reaction zone. 
At projectile energies below the free pion production threshold 
($E^{th({\pi^o})}_{NN}$=280 MeV) pions and hard photons 
($E_\gamma \geq 30$ MeV) carry information on the initial compressional 
and high density phase of the heavy-ion collision which might evolve towards 
a hot thermalized nuclear system \cite{CassingReport,Schu97}.

In first order, subthreshold pions and photons are created following either 
single or several nucleon-nucleon collisions, the nuclear medium providing 
the additional energy needed to surpass the production threshold. 
Therefore subthreshold pions and photons can be viewed as a snapshot of 
the dynamical phase-space occupancy of participant nucleons at the early 
instant of the collision. 
However, extracting the information relevant to the thermodynamical state 
of the system is far from being straightforward. 
One must rely on models which describe the complicated many-body problem 
of a heavy-ion collision.
Alternatively, a purely empirical approach consists in the comparison 
of various observables between heavy-ion collisions and proton-nucleus 
collisions. 
Since a compressional phase is very unlikely to occur in the latter, 
the in-medium nucleon-nucleon collisions, i.e. the longitudinal
momentum dissipation, can be studied without the complication due to 
the dynamics and thermodynamics of heavy-ion collisions.

For the present measurement the proton beam energy was chosen to be 190 MeV, resulting from a compromise 
between the concurrent requirements to accumulate a usable set of data 
and to be well below the pion production threshold.
The energy spectra, up to the maximum energy available in the reaction 
for the production of a single particle, were measured together with 
their angular distributions.
The target mass dependence of the production cross sections,
studied as a function of the energy of the produced particle, is exploited to 
characterize the production mechanism and, in the case of pions, 
their propagation properties through nuclear matter.
It is concluded that the production mechanism of subthreshold particles 
changes with increasing particle energy from a production 
in individual nucleon-nucleon collisions 
to mechanisms involving more than two nucleons. 
Although several similar studies have been published earlier [3-8],
the information collected in the present work is unique as it combines for the
first time the measurement of complete energy spectra with the concurrent 
observation of strongly interacting particles, pions, 
and non-interacting particles, photons.

\section{Experimental setup}
\label{sec:experiment}

The 190 MeV proton beam was delivered by the superconducting
cyclotron AGOR at KVI. 
Natural C, Ca, Ni and W targets, respectively 18.1, 11.2, 6.7
and 5.0 mg/cm$^2$ thick, were irradiated with 10 to 25 nA beams in bunches
of 2 ns at a rate of 60 MHz. 
Within these conditions the interaction rate 
was respectively for the four targets 0.23, 0.13, 0.09, and 0.05 nuclear 
interactions per beam pulse. 
These values set the counting rate in an individual photon detection module to
1 kHz and induced negligible double hits and random coincidences. 
Direct photons
and decay photons from neutral pions ($\pi^{o}\rightarrow \gamma \gamma$,
BR=98.8\%) were detected with the photon spectrometer TAPS
configured in six blocks of 64 BaF$_2$ modules, each module being associated 
with a charged-particle-veto detector. 
The blocks were positioned at the most forward angles, allowed by technical 
constraints, on both sides of the beam. 
They covered a quasi-continuous polar angle range above
$\theta=60^{o}$ and an azimuthal angle range of 
$\pm 20^{o}$, hence covering approximately 15\% of $4\pi$.
The distance from the target to the active front face of the blocks 
($d=66$ cm) offered a flight path large enough to discriminate photon hits 
against baryon (mainly neutron and proton) hits through their time-of-flight 
measurement. 
The energy deposited in the BaF$_2$ modules was calibrated
and the gain continuously monitored during the experiment using the 38.5 MeV
peak energy-deposition of cosmic muons.  
Time-of-flight was calibrated using the prompt photon signal and the known 
time interval between proton beam pulses (16.7 ns).
To enrich the recorded events with photons and neutral pions the trigger 
electronics was set up to select events in which at least one, 
respectively two, neutral hits (defined as a hit in a BaF$_2$ module 
and no hit in the charged-particle-veto detector) were present, depositing at least 
15 MeV equivalent photon energy.
A downscaled trigger consisting of a neutral hit of at least 0.4 MeV served 
as minimum-bias trigger. 
An exhaustive description of the TAPS detector, its associated electronics 
and trigger scheme can be found in \cite{Schu97}.


In the offline analysis photon hits were identified by combining the
information delivered by the BaF$_2$ modules and by exploiting the
characteristic pulse shape and time of flight of photons.  The
direction and energy of the impinging photon was derived from the
analysis of the electromagnetic-shower
topology\cite{Martinez97nim,nimmig} which provided also additional
discrimination criteria against baryons and cosmic muons.  
Neutral pions were identified, with a full width at half maximum of 10\%, through
an invariant mass analysis of identified photon pairs.  In the
resulting invariant mass spectrum combinatorial background can be safely neglected
because of the weak $\gamma$ multiplicity at the considered 
bombarding energies ($M_{\gamma}\approx 10^{-3}$).
Photon pairs with an invariant mass between 100 and 160 MeV were
adopted as stemming from the decay of the produced neutral pions. 
Close to one million pions were detected and identified for each target.

\begin{figure}
\begin{center}
\includegraphics[width=7cm]{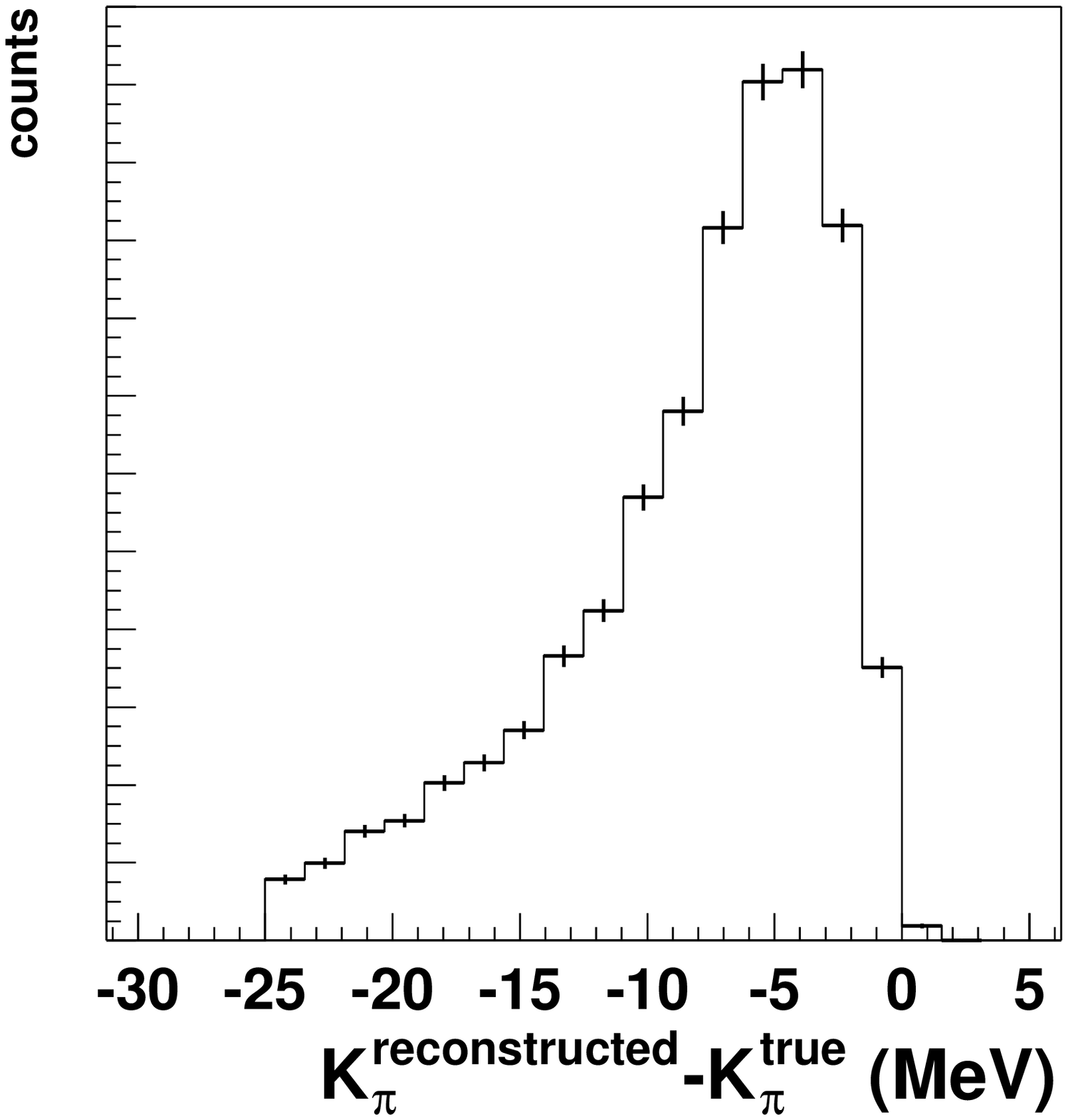}\includegraphics[width=7cm]{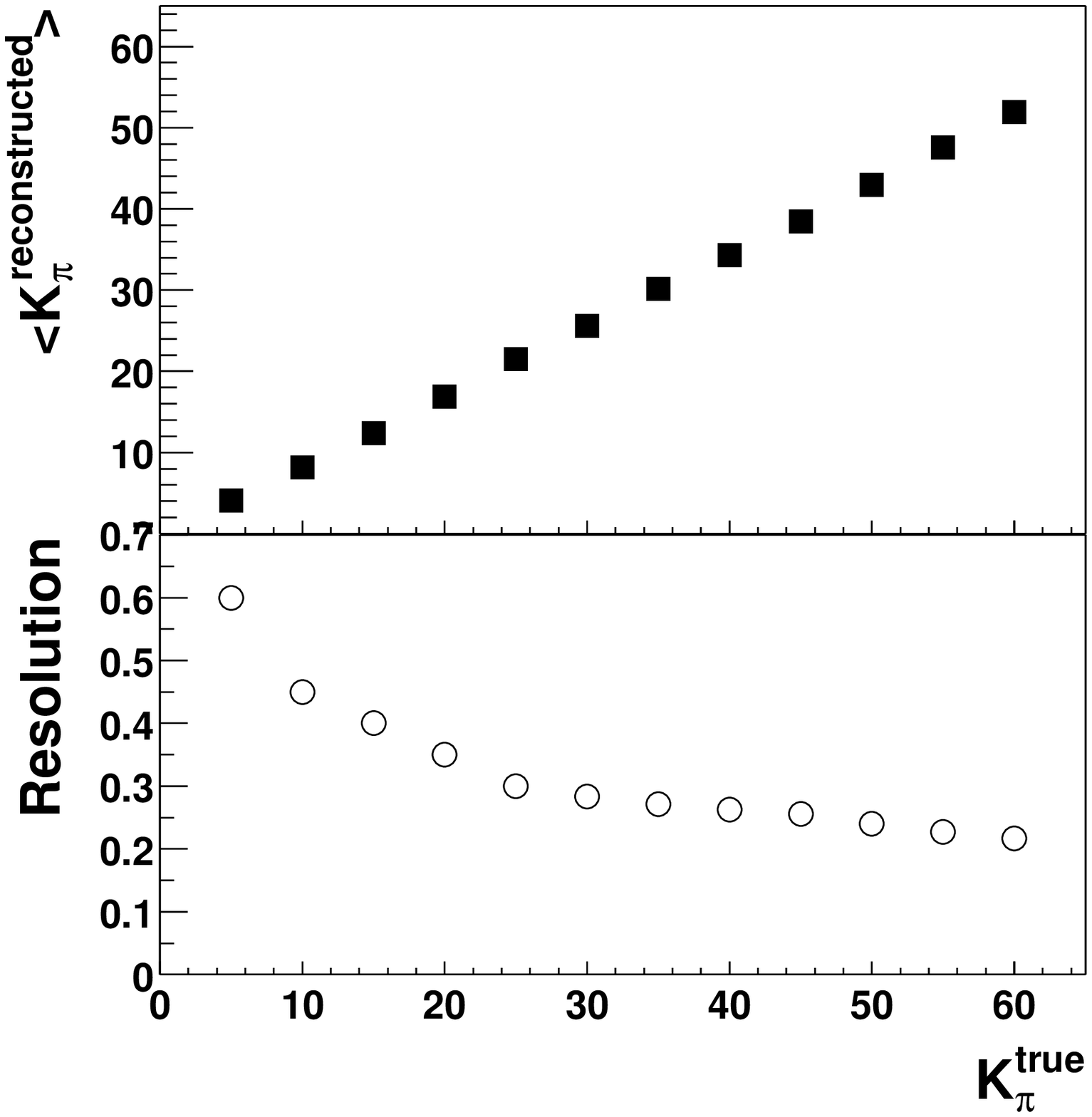}
\caption{Left: typical simulated response function of TAPS to mono-energetic neutral pions of 25 MeV, represented by the difference between the true initial pion kinetic energy and the reconstructed pion kinetic energy modified by the detection system and the reconstruction altogether. Right: Performance of the neutral pion momentum reconstruction method, as a function of the neutral pion kinetic energy (top: mean of the reconstructed energy, bottom: resolution).}
\label{fig:response function}
\end{center}
\end{figure}

For every identified neutral pion, the 4-momentum was calculated 
through a constrained minimization method using the 4-momentum of the 
two photons. 
This special energy reconstruction method was developed\cite{Aphecetche98} 
because the standard technique\cite{Baer81} leads to unphysical results\cite{Korzecka2000} 
(pion energies above the kinematical limit) in our
case, where the pions i) are of very low energy and ii) take away a
large (up to 100\%) part of the total energy available in the proton+nucleus system. 
Performances of this new method have been carefully checked
using GEANT simulations\cite{KANE96} of mono-energetic neutral
pions. Unlike the standard method, our minimization method leads to a
non-symmetric response function (Fig. \ref{fig:response function}),
with a reconstructed mean energy which is about 88\% of the true value, 
and an energy resolution depending on pion energy
(Fig. \ref{fig:response function}).


The neutral pion efficiency was
calculated with the help of GEANT simulations taking as input an extrapolation
$D(K_{\pi^o},\Omega_{\pi^o})$ to the full solid angle of the measured double
differential cross-sections $d^{2}\sigma /d\Omega _{\pi^o }dK_{\pi^o }$.
$D(K_{\pi^o},\Omega_{\pi^o})_{lab}= d\sigma/dK_{\pi^o}$ 
(angular distribution has been chosen isotropic, in agreement with the data)
where $d\sigma/dK_{\pi^o}$ was obtained through the following iterative procedure.
The distribution of the measured kinetic energy, $K_{\pi^o}$, was parametrized as :

\begin{equation}
\label{eq:kpi distribution}
\frac{d\sigma}{dK_{\pi}} \propto K_{\pi^o}^{\alpha} \times 
\exp\left( - \frac{K_{\pi^o}}{T_1} \right) \times \frac{1}{1+\exp\left(
\frac{K_{\pi^o}-K_0}{T_2}\right)}
\end{equation}

Expression \ref{eq:kpi distribution} was selected as the initial guess of 
the iteration procedure and fed into the simulation.
The parameters $\alpha,T_1,T_2,K_0$ were then tuned until the output of 
the simulation reproduces the measured distribution.

The photon efficiency was calculated using the same scheme, 
with the following input distribution :
$$\left(\frac{d^2\sigma_\gamma}
   {dE_\gamma\Omega_\gamma}\right)_{NNcm}^{E_\gamma>40 MeV} \propto 
   \exp\left(\frac{E_\gamma-E_0}{\sigma_E}\right) \times 
      \left[ \alpha_1\sin^2(\theta_\gamma)+\alpha_2 \right]$$

The resulting pion and photon efficiencies are reported in Table \ref{tab1}.



\section{Cross-sections}

The neutral pion total cross sections were calculated as:

$$\sigma_{\pi^o}^{4\pi}={\mathcal{C}_D}\times
N_{\pi^o}^{raw}/\epsilon_{\pi^o}^{4\pi}$$

where $N_{\pi^o}^{raw}$ is the raw
number of identified pions, $\mathcal{C}_D$ is a normalization factor
including target thickness, beam intensities and trigger conditions, and
$\epsilon_{\pi^o}^{4\pi}$ is the global efficiency. 
The resulting cross-sections $\sigma_{\pi^o}^{4\pi}$ are reported in 
Table \ref{tab1}.

\begin{figure}
\begin{center}
\includegraphics[width=7cm]{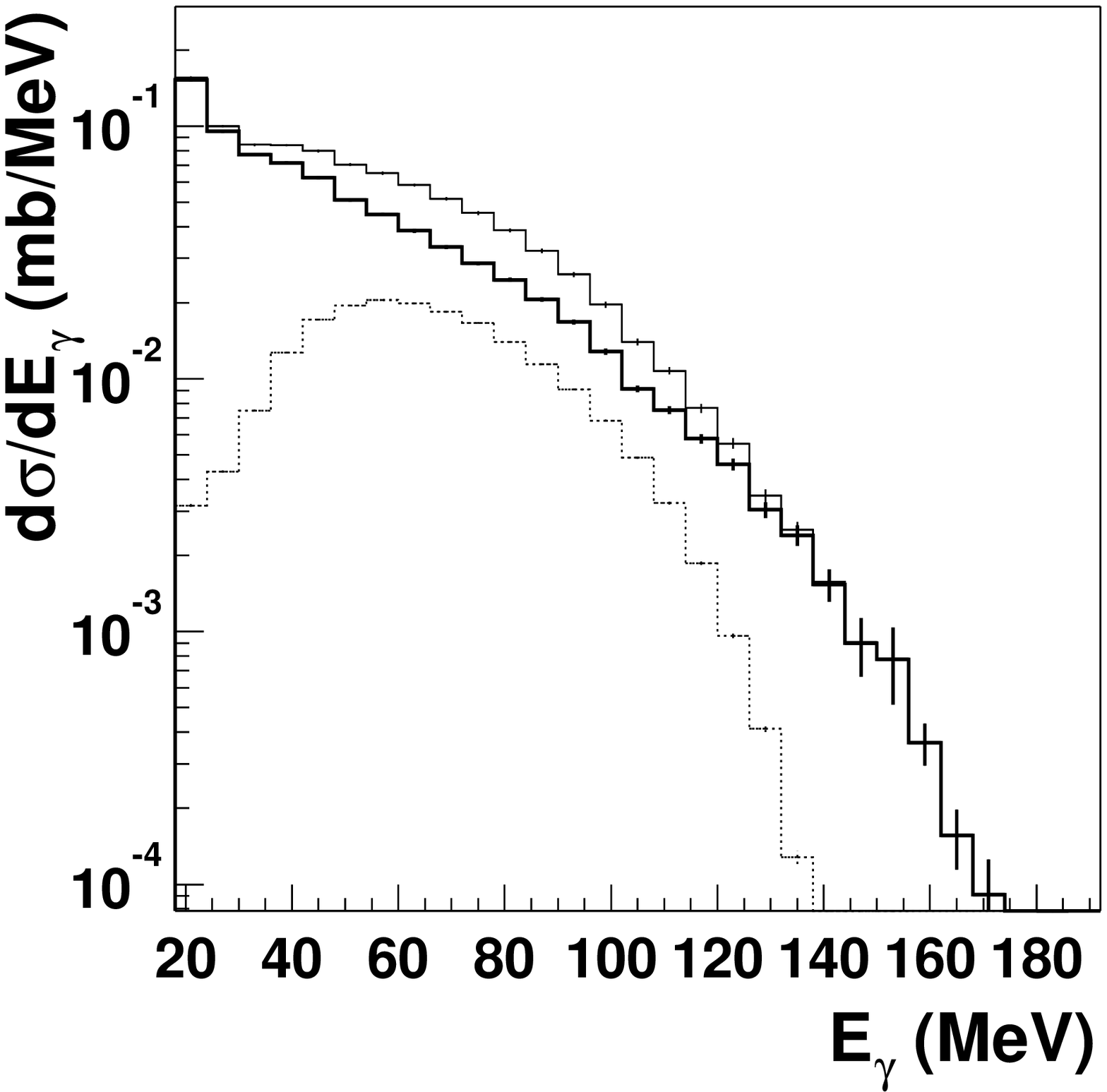}\includegraphics[width=7cm]{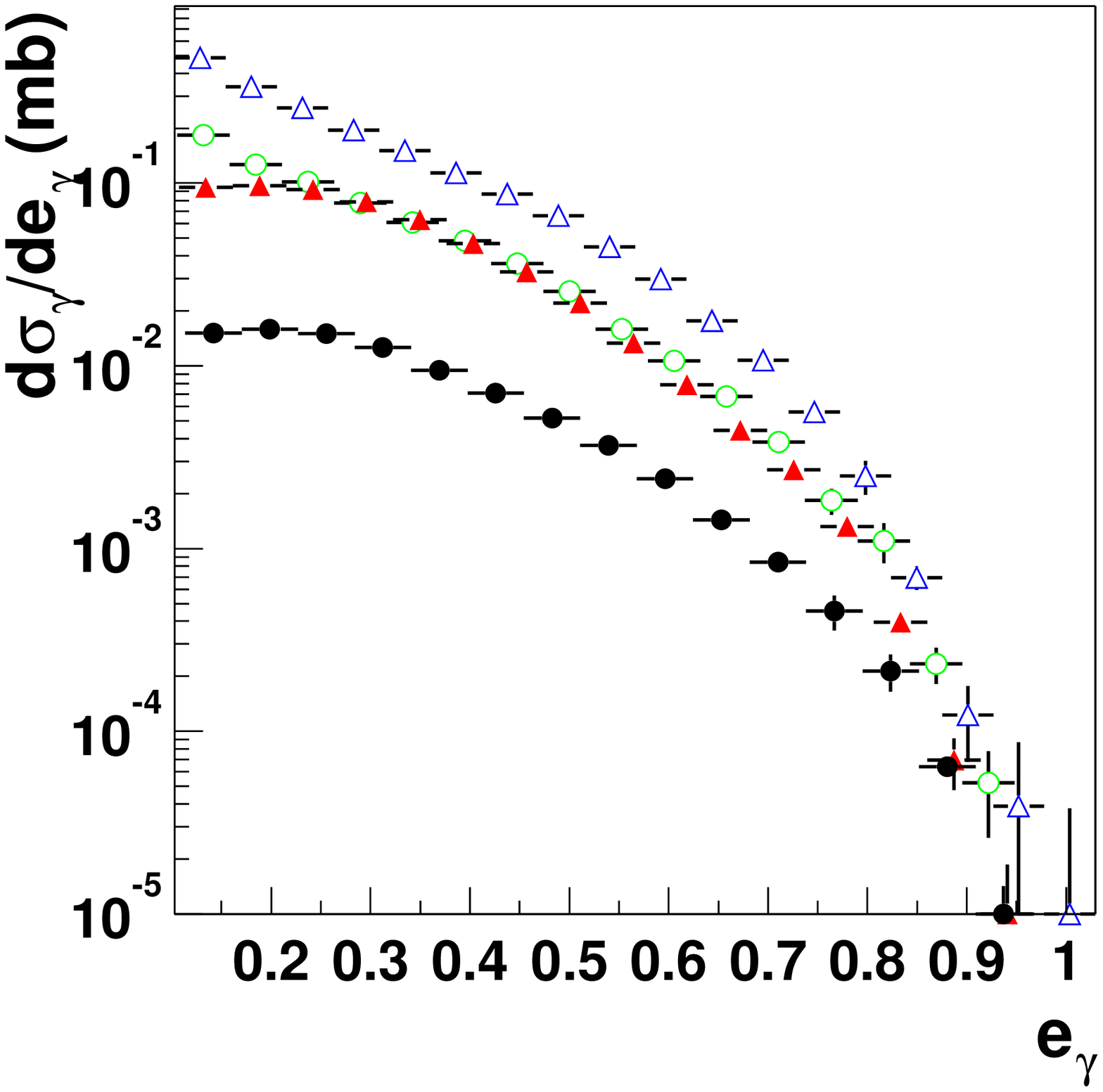}
\caption{Left: contribution of decay photons from $\pi^0$ (dotted line) 
in the total (thin solid line) photon spectra, 
and the resulting direct photon spectra (bold solid line). 
Right: direct photon energy spectra measured in p+A reactions at 190 MeV, 
as a function of the reduced photon energy (from bottom to top:
C, Ca, Ni and W).}
\label{fig:dsigma_photon}
\end{center}
\end{figure}

To obtain direct photon cross-sections an additional treatment is required, 
as the contribution of the decay photons from neutral pions must be subtracted 
from the total photon cross-section. The latter is obtained as
$$\sigma_\gamma=
{\mathcal{C}_D^\prime} \times N_\gamma^{raw} / \epsilon_\gamma^{4\pi}$$
where $\epsilon_\gamma^{4\pi}=12\%$  is the efficiency for photons and 
$N_\gamma^{raw}$ is the raw number of identified photons.
The contribution of the decay photons was estimated with the help of GEANT 
simulations taking as input the extrapolated double-differential cross-section 
$D(K_{\pi^o},\Omega_{\pi^o})$. The contribution of decay photons amounts to about 30\% of the photon spectrum (Fig. \ref{fig:dsigma_photon}).

From the photon cross section the probability $P_\gamma$ to produce a single photon
in an individual proton-neutron collision was evaluated according
to the prescription of \cite{Clayton92b}.
We observe a good agreement of this probability with the systematics 
presented in \cite{Clayton92b} (averaged for Ca, Ni, W targets, the $P_\gamma$ equals to $(7.4\pm1.4)\times 10^{-4}$, compared to $9.4\times 10^{-4}$ expected from the systematics) with the exception
of the C target (factor 2 below the systematics).
The neutral pion cross sections can be compared to the results of 
\cite{Bellini89} obtained with proton beams of 200 MeV kinetic energy.
While we observe the same (within error bars) mass scaling law of 
the total cross section for neutral pion production, the cross section
values reported in \cite{Bellini89} are higher by a factor of approximately 3.
Such a strong rise can not be attributed to the 10 MeV difference
in beam energies.
The problem of inconsistencies in the cross section for pion
production in proton-nucleus reactions is discussed in detail 
in \cite{Wilschut01}.


Thanks to the good statistics accumulated, errors in this experiment 
are mainly systematic in nature, and amount to 10\% for $\mathcal{C}_D$, 
7\% for $\epsilon_{\pi^o}^{4\pi}$, 20\% for $\epsilon_{\gamma}^{4\pi}$, 
leading to total relative errors
of about 32\% for the direct photon cross sections and 12\% for the total pion
cross sections.



\begin{figure}
\begin{center}
\includegraphics[width=7cm]{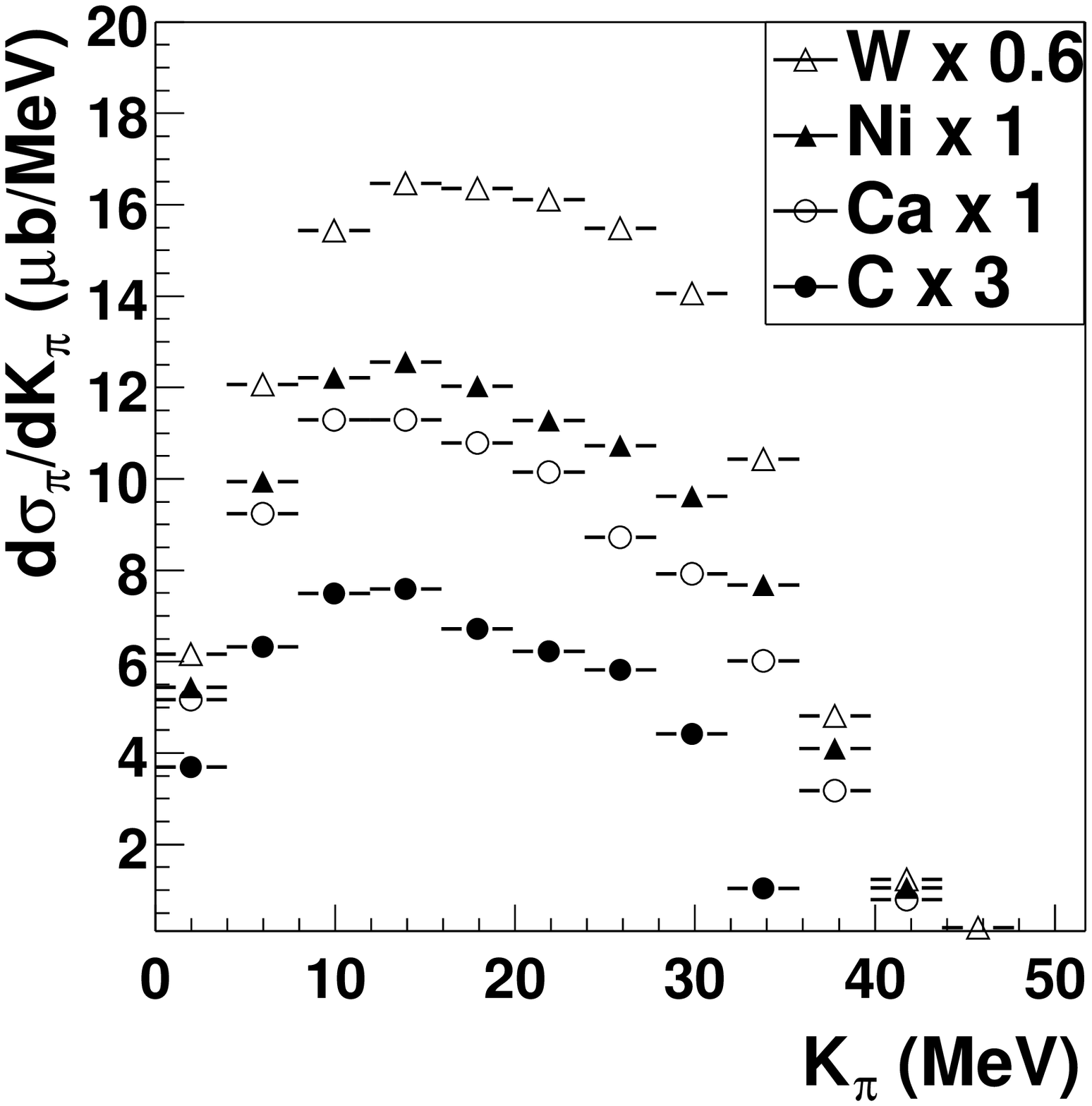}\includegraphics[width=7cm]{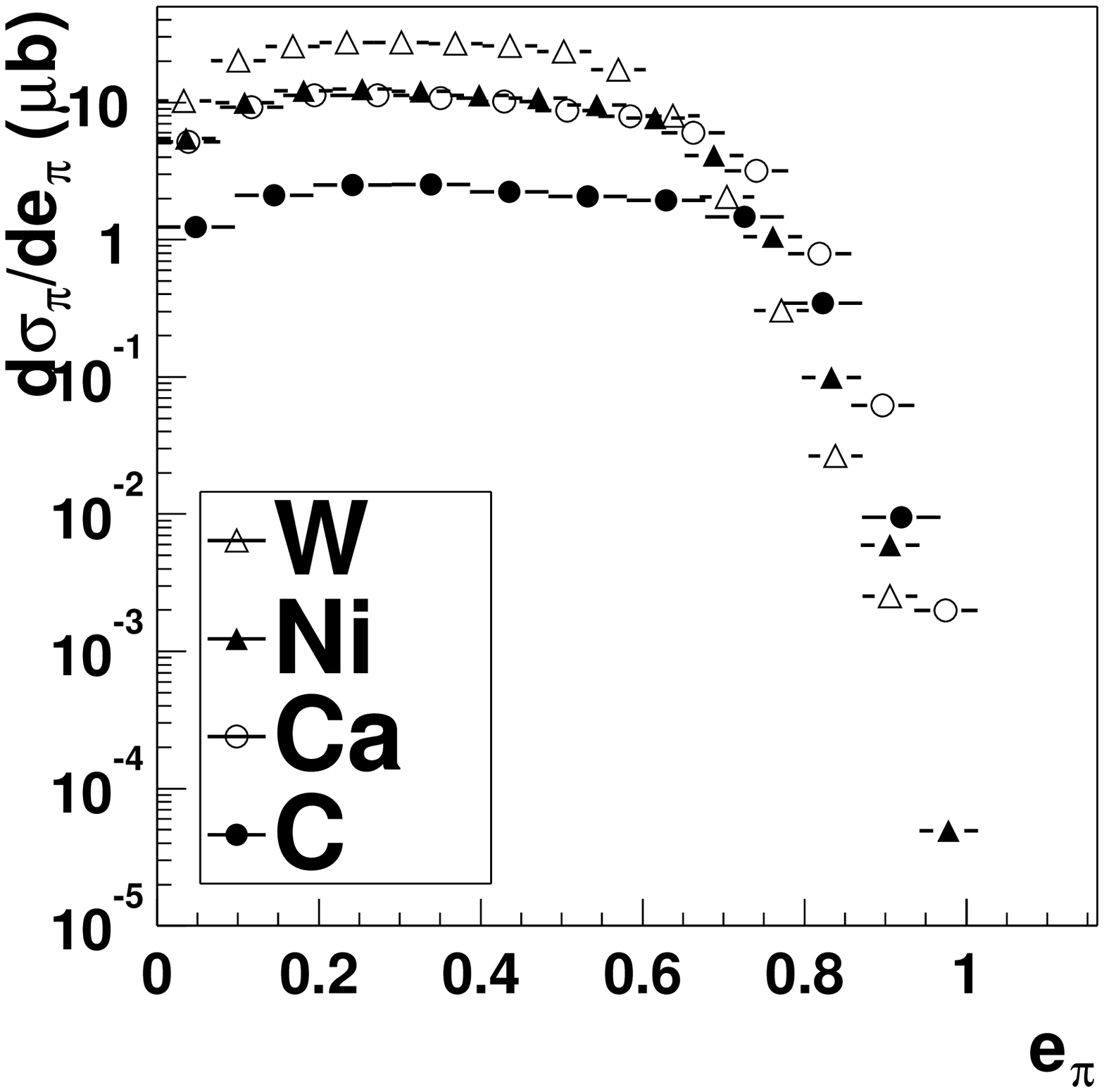}
\caption{Neutral pion energy spectra measured in p+A reactions at 190 MeV, 
as a function of kinetic pion energy (left) and of the reduced neutral pion energy 
$e_\pi=K_{\pi}/K^{max}$ (right, see text).}
\label{fig:dsigma_pi}
\end{center}
\end{figure}

The energy spectra of particles (Figs. \ref{fig:dsigma_pi} 
and \ref{fig:dsigma_photon}) produced from the four different targets 
are conveniently compared if one defines a reduced energy, $e_{r}$,
which represents the fraction of the available kinetic-energy, $K^{max}$, 
carried away by the particle.
This energy is calculated (Table \ref{tab2}) assuming that the reaction 
proceeds through the fusion of the proton and the target nucleus:  
$K^{max}=\sqrt{s}-\left( A+1\right) M_{N}-Q$, 
where $\sqrt{s}$ represents the total energy available in the center-of-mass,
$M_N$ the nucleon mass and $Q$ the fusion reaction Q-value.
For photons $K\equiv E_{\gamma }$. 
Within this representation (Figs. \ref{fig:dsigma_pi} and 
\ref{fig:dsigma_photon}, right panels) particles with $e_{r}=1$ take away 
all the available energy and the mechanism would be pionic fusion\cite{Horn96} 
for pions or radiative capture for photons.

\section{Analysis of target mass dependence}

The dependence with the target mass $A$ of the pion and photon cross-sections
can be described by a power law, $d\sigma /dx\propto A^{\alpha(x)}$ ,
where $x$ is either the polar angle of the emitted particle or its reduced 
kinetic energy. 
This widely used parametrization at relativistic
\cite{Elmer96,Wolf98,Miskowiec94,Barth97} as well as ultrarelativistic
\cite{Albrecht98,Aggarwal98} energies for a variety of produced particles 
gives valuable indications on the particle-production mechanism. 
Since one is interested in the primordial particle production, 
one must first understand how the primordial production is modified 
by subsequent absorption or re-scattering during the propagation through 
the nuclear medium. 
This has been done by examining the target mass dependence of the differential 
cross-section, $d\sigma /d\theta _{\pi^o}$, in terms of an elementary 
geometrical model which tracks the pion through the nucleus after its creation 
\cite{Aphecetche98}.
Within this picture our data are best described if one assumes that:
i) primordial pions are produced in a surface shell of the nucleus
with thickness $\lambda_p=2.4$ fm, which is equal to 
the proton mean free path, and
ii) the pion mean free path, independent of the pion energy, is equal to
$\lambda_\pi=6$ fm in good agreement with the values established elsewhere
\cite{Mayer93}.
The first assumption implies that pions detected in the backward hemisphere 
are quasi unaffected by re-interaction in the medium and those pions can 
thus be considered as primordial pions. 

\begin{figure}
\begin{center}
\includegraphics[width=9cm]{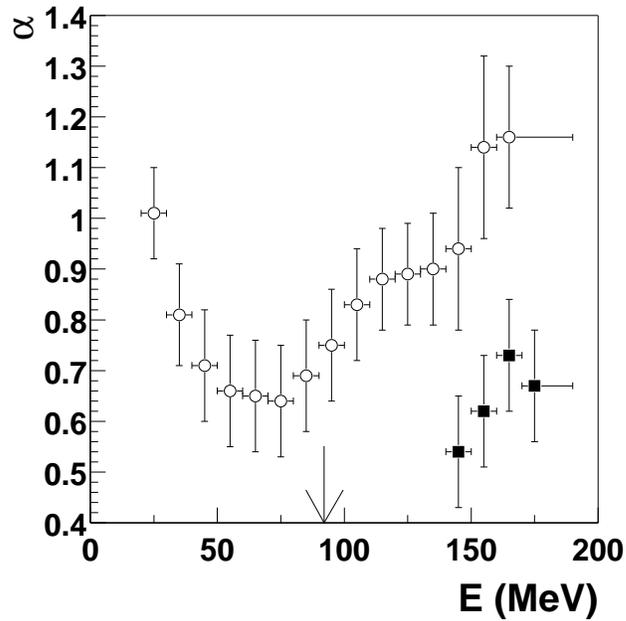}
\caption{Evolution of the $\alpha$ parameter as a function of the particle 
energy, for photons (open symbols) and neutral pions (closed symbols). 
The arrow indicates the separation between below- and above-threshold photons.}
\label{fig:alpha}
\end{center}
\end{figure}

The target mass $A$ dependence of neutral pion and photon cross-sections was analyzed
in terms of the power-law parameter $\alpha$.
The carbon data have been excluded from this analysis because of its 
extraordinary behavior (see also \cite{Goethem98}).
All neutral pions and photons with energy above 92 MeV were produced
below the free NN threshold.
For hard photons produced below threshold the $\alpha$ parameter
(Fig. \ref{fig:alpha}) continously increases from $\alpha=2/3$ 
to values close to or even above one.
This behavior is interpreted as an evolution of the particle-production
mechanism from a surface production for the least energetic particles towards
a volume production for the most energetic particles. 
In other words, the observed $\alpha$ dependence witnesses a transition from 
particles produced in first-chance nucleon-nucleon collisions to particles 
produced in multiple successive binary collisions. 
A possible mechanism for the latter is the process of photoabsorption of
subthreshold pions in the nuclear medium ($NN\rightarrow NN\pi$ 
and subsequently $N\pi\rightarrow N\gamma$), as proposed in \cite{Gudima96}.
The $\alpha$ dependence with the photon energy (Fig.\ref{fig:alpha}) 
indicates another strong rise when the energy of photons decreases. 
This can be understood invoking again the participation of all nucleons 
with the difference that in this case there is enough energy available, not only 
in first chance collisions but also in secondary collisions. 
A detailed analysis of this photon energy region was performed by another 
TAPS experiment during the same experimental campaign at KVI \cite{Goethem98}.

In the case of neutral pions such firm  conclusions can not be drawn from
the available experimental data.
Within error bars, a rise of the $\alpha$ parameter towards unity 
or a constant value of $\alpha=2/3$ can not be distinguished.
It must be emphasized that, compared to photons, pions are less adequate probes
for a model-independent analysis.
Their production often involves baryonic resonances as intermediate
steps \cite{Matulewicz00}, and 
the energy dependence of the power-law parameter $\alpha$ might
be affected by the energy variation of the pion absorption length,
 an effect not accounted for in the simple geometrical model used here.

\section{Conclusion}

A complete data set on the production of neutral pions
and photons in reactions of 190 MeV protons with targets ranging in mass
from $A=12$ to $A=184$ has been obtained.
Energy spectra up to the kinematical limit have been presented. 
A simple geometrical model, assuming neutral pion production in a surface
shell of the target nucleus and subsequent pion propagation
with the mean free path equal to $\lambda _{\pi }=6$ fm, agrees well
with the observed angular distribution of neutral pions.
The analysis of the mass scaling law of subthreshold particle cross-sections supports
the mechanism of particle production in first-chance nucleon-nucleon 
collisions, except in the case of most energetic particles 
(up to the kinematical limit), where the role of multi-step 
processes becomes evident.

The present data furnish new and valuable information for tuning unknown parameters
 of dynamical models for heavy-ion collisions\cite{CassingReport}, like in-medium cross sections or 
 the implementation of the Pauli principle. Further evidence for new production mechanisms like
 $pn\rightarrow d\pi \left( \gamma \right)$ or 
$\pi N\rightarrow N\gamma$ \cite{Gudima96} has been given. The complicated and hardly understood
 production mechanism of hadronic particles, 
like pions and kaons \cite{Koptev01},
affected by reabsorption and rescattering, stresses the importance
of hard-photon measurements as presented in this letter. 
Once hadronic and electromagnetic probes are well understood, the dynamical transport model codes may
 provide a better description of the particle production in heavy-ion collisions, which is the base to
 unravel the thermodynamic properties of nuclear matter.


\section*{Acknowledgements}
We would like to thank the AGOR team for providing
 a high quality proton beam.
This work was in part supported
by the Institut National de Physique Nucl\'eaire et de Physique
des Particules, France,
by the Dutch Stichting voor Fundamenteel Onderzoek der Materie, The Netherlands,
and by the Grant Agency of the Czech Republic.



\bibliography{pxx190}



\newpage

\begin{table}
\begin{center}
\caption{
Cross sections of neutral pions and photons measured
 in p+A reactions at 190 MeV. $N_{\pi^o}^{raw}$ is the number of neutral pions recorded, $\mathcal{C}_D$ takes into account 
   target thickness, beam intensity and trigger rates.    
   $\sigma_{\pi^o}^{4\pi}={\mathcal{C}_D}\times N_{\pi^o}^{raw}/\epsilon_{\pi^o}^{4\pi}$ 
   is the total neutral pion cross-section, calculated using
   estimated neutral pion global efficiency.    
   The direct photon cross-section 
   $\sigma_\gamma^{direct}=\sigma_\gamma^{total}-\sigma_{\gamma}^{\pi^o \rightarrow \gamma \gamma}$ 
   is then deduced from the total photon
   cross-section
   by subtracting contribution from decay photons. All
   photon cross-sections are given for $E_\gamma^{lab} \geq 40$ MeV.
}
\label{tab1}
\begin{tabular}{|c|c|c|c|c|}
\hline
~ & C & Ca & Ni & W\\
\hline
${\mathcal{C}_D}\times 10^{6}$ ($\mu$b)  & $1.45\pm0.15$ & $10.2\pm1.0$ & $2.38\pm0.24$ & $12.5\pm1.2$ \\
\hline
$N_{\pi^o}^{raw}\times 10^{-5}$  & $3.042\pm0.005$ & $2.017\pm0.004$ & $9.870\pm0.01$ & $4.166\pm0.006$ \\
\hline
$\epsilon_{\pi^o}^{4\pi}$ (\%) & $2.65\pm0.18$  & $2.43\pm0.18$ & $2.43\pm0.18$ & $2.44\pm0.18$ \\
\hline
$\sigma_{\pi^o}^{4\pi}$ ($\mu$b)  &  $16\pm2$ & $85\pm10$ & $97\pm12$ & $215\pm26$ \\
\hline
\hline
$\epsilon_{\gamma}^{4\pi}$ (\%) & $0.12\pm0.02$  & $0.12\pm0.02$ & $0.12\pm0.02$ & $0.12\pm0.02$ \\
\hline
$\sigma_{\gamma}^{total}$ ($\mu$b) & $88\pm20$ & $520\pm120$ & $567\pm130$ & $1383\pm320$ \\
\hline
$\sigma_\gamma^{\pi^{o} \rightarrow \gamma \gamma}$ ($\mu$b) & $28\pm3$ & $149\pm18$ & $170\pm20$ & $376\pm45$ \\
\hline
$\sigma_{\gamma}^{direct}$ ($\mu$b) & $60\pm20$ & $372\pm120$ & $397\pm130$ & $1007\pm310$ \\
\hline
\end{tabular}
\end{center}
\end{table}


\begin{table}
\begin{center}
\caption{Reaction Q-values (from \protect\cite{NUBASE97}), used to compute the maximum available $K_{max}$ in A(p,X) reactions : $Q_{gg}=\Delta M_{A+1}-\Delta M_{A}-\Delta M_1 + m_X$ where $X=\pi^o,\gamma$. }
\label{tab2}
\begin{tabular}{|c|c|c|}
\hline
Reaction & $Q_{gg}$ (MeV) & $K_{max}$ (MeV)\\
\hline
$^{12}_{6}$C$_{g.s.}(p,\pi^0)^{13}_{7}$N$_{g.s.}$ & 133.0 & 41 \\
\hline
$^{40}_{20}$Ca$_{g.s.}(p,\pi^0)^{41}_{21}$Sc$_{g.s.}$ & 133.9 & 51 \\
\hline
$^{58}_{28}$Ni$_{g.s.}(p,\pi^0)^{59}_{29}$Cu$_{g.s.}$ & 131.5 & 55 \\
\hline
$^{184}_{74}$W$_{g.s.}(p,\pi^0)^{185}_{75}$Re$_{g.s.}$ & 129.6 & 59 \\
\hline
\hline
$^{12}_{6}$C$_{g.s}(p,\gamma)^{13}_{7}$N$_{g.s.}$ & -1.94 & 176 \\
\hline
$^{40}_{20}$Ca$_{g.s.}(p,\gamma)^{41}_{21}$Se$_{g.s.}$ & -1.08 & 186 \\
\hline
$^{58}_{28}$Ni$_{g.s.}(p,\gamma)^{59}_{29}$Cu$_{g.s.}$ & -3.42 & 190 \\
\hline
$^{184}_{74}$W$_{g.s.}(p,\gamma)^{185}_{75}$Re$_{g.s.}$ & -5.40 & 194 \\
\hline
\end{tabular} 
\end{center}
\end{table}

\newpage

\end{document}